\documentclass[aip,amsmath,amssymb, reprint]{revtex4-1}

\usepackage{graphicx}% Include figure files
\usepackage{dcolumn}% Align table columns on decimal point
\usepackage[utf8]{inputenc}
\usepackage[T1]{fontenc}
\usepackage{color}
\usepackage{latexsym}
\usepackage{siunitx}
\usepackage{amsmath}
\usepackage{upgreek}
\usepackage{natbib}
\usepackage[normalem]{ulem}
\useunder{\uline}{\ul}{}

\begin{document}

\preprint{AIP/123-QED}

\title{Comparison of Dielectric Loss in Titanium Nitride and Aluminum Superconducting Resonators}
% Force line breaks with \\

\author{A. Melville}
\thanks{alexander.melville@ll.mit.edu; These authors contributed equally to this work.}
\affiliation{MIT Lincoln Laboratory, 244 Wood Street, Lexington, MA 02421, USA}
\author{G. Calusine}
\thanks{alexander.melville@ll.mit.edu; These authors contributed equally to this work.}
\affiliation{MIT Lincoln Laboratory, 244 Wood Street, Lexington, MA 02421, USA}
\author{W. Woods}
\thanks{alexander.melville@ll.mit.edu; These authors contributed equally to this work.}
\affiliation{MIT Lincoln Laboratory, 244 Wood Street, Lexington, MA 02421, USA}
\author{K. Serniak}
\affiliation{MIT Lincoln Laboratory, 244 Wood Street, Lexington, MA 02421, USA}
\author{E. Golden}
\affiliation{MIT Lincoln Laboratory, 244 Wood Street, Lexington, MA 02421, USA}
\author{B. M. Niedzielski}
\affiliation{MIT Lincoln Laboratory, 244 Wood Street, Lexington, MA 02421, USA}
\author{D. K. Kim}
\affiliation{MIT Lincoln Laboratory, 244 Wood Street, Lexington, MA 02421, USA}
\author{A. Sevi}
\affiliation{MIT Lincoln Laboratory, 244 Wood Street, Lexington, MA 02421, USA}
\author{J. L. Yoder}
\affiliation{MIT Lincoln Laboratory, 244 Wood Street, Lexington, MA 02421, USA}
\author{E. A. Dauler}
\affiliation{MIT Lincoln Laboratory, 244 Wood Street, Lexington, MA 02421, USA}
\author{W. D. Oliver}
\affiliation{MIT Lincoln Laboratory, 244 Wood Street, Lexington, MA 02421, USA}
\affiliation{Research Laboratory of Electronics, Massachusetts Institute of Technology, Cambridge, MA 02139, USA}
\affiliation{Department of Electrical Engineering \& Computer Science, Massachusetts Institute of Technology, Cambridge, MA 02139, USA}

\date{\today}% It is always \today, today,

\begin{abstract}
Lossy dielectrics are a significant source of decoherence in superconducting quantum circuits. In this report, we model and compare the dielectric loss in bulk and interfacial dielectrics in titanium nitride (TiN) and aluminum (Al) superconducting coplanar waveguide (CPW) resonators. We fabricate isotropically trenched resonators to produce a series of device geometries that accentuate a specific dielectric region's contribution to resonator quality factor. While each dielectric region contributes significantly to loss in TiN devices, the metal-air interface dominates the loss in the Al devices. Furthermore, we evaluate the quality factor of each TiN resonator geometry with and without a post-process hydrofluoric (HF) etch, and find that it reduced losses from the substrate-air interface, thereby improving the quality factor. 
\end{abstract}

\maketitle

% Paragraph 1

Dielectric loss from material interfaces limit performance in superconducting quantum devices.\cite{Martinis2005, Gao2006, Gao2008, OConnell2008, Barends2008, Vissers2010, Weber2011, Wenner2011, Chang2013, Woods2019} The magnitude of dielectric loss at these interfaces is determined by the materials and processes used to fabricate the devices. As such, it is imperative to develop a quantitative framework to understand how the loss at interfaces is affected by the choice of superconducting metal and subsequent fabrication steps. Significant work has focused on identifying which regions of a device may most strongly limit performance by modeling their electric field participation.\cite{Wenner2011, Wang2015, Gambetta2017} Separately, many reports have compared the performance of devices constructed using different materials or fabrication processes.\cite{Barends2007, Wang2009, Sage2011, Megrant2012, Bruno2015, Place2020,Tsioutsios2020} By combining these ideas, differences in quality factor can be directly attributed to loss in specific dielectric regions, enabling further data-driven improvements to device performance.

% Paragraph 1.5

In this work, we use the surface-loss extraction (SLE) process, outlined in Ref$.$ \onlinecite{Woods2019}, to model the dielectric regions of TiN and Al superconducting resonators and calculate the loss tangents of these regions based on the measured quality factors. We find that the metal-air interface of the Al resonators is an order of magnitude more lossy than for TiN resonators. We also used a post-process HF etch to reduce the loss in the TiN devices and applied the SLE process to attribute the reduction specifically to the substrate-air interface.

% Paragraph 2

\begin{figure}
\includegraphics[scale=1]{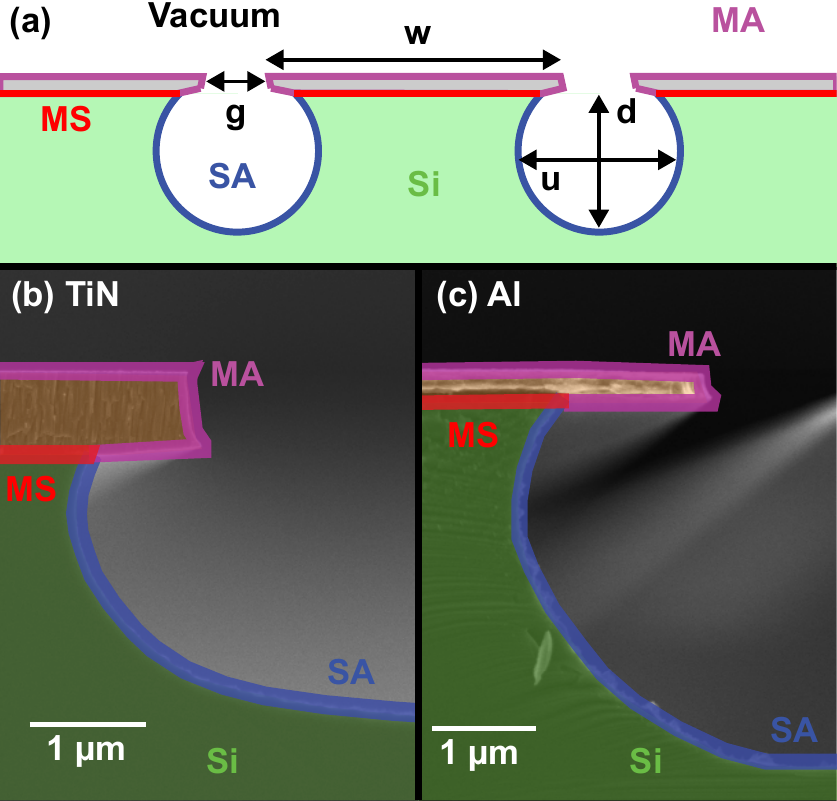}
\caption{\label{fig1} (a) Diagram illustrating a cross-section of a typical coplanar waveguide resonator with width (w), gap (g), trench depth (d) and degree of undercutting (u). The dielectric regions are labeled as follows: MS = Metal-Substrate interface, SA = Substrate-Air interface, MA = Metal-Air interface, Si = Silicon substrate. Representative cross-section scanning electron micrographs (SEM) of deeply isotropically trenched TiN (b) and Al (c) resonators. The cross-sections are used to generate a finite element model for calculating the participation ratio of each dielectric region.\cite{COMSOL}}
\end{figure}

We differentiate four dielectric regions from which we can extract a loss tangent (see Fig$.$ \ref{fig1}): the metal-substrate interface (MS), the substrate-air interface (SA), the metal-air interface (MA), and the silicon substrate~(Si). Two-level-system (TLS) defects in these dielectric regions limit the quality factor of a resonator to 
\begin{equation} \label{eqn1}
Q_{\mathrm{TLS}}^{-1} = \sum_{r} p_{r} \tan \delta_{r} \
\end{equation}
where $p_{r}$ is the geometry-dependent electric field participation ratio and $\tan \delta_{r}$ is the loss tangent of dielectric region $r$. By measuring the quality factor, $Q$, of a set of four specific resonator geometries with a distinct distribution of participation values,\cite{Supplemental} we numerically solved for the loss factor of each dielectric region,\cite{Woods2019} which we then convert to a loss tangent using a reasonable set of assumptions about the dielectric constant and layer thickness.\cite{Assumptions} The four geometries were each designed to accentuate participation in one of the four dielectric regions relative to the others. Therefore, we refer to the geometries as ``MS design,'' ``SA design,'' ``MA design,'' and ``Si design,'' corresponding to the dielectric region being emphasized. All geometries were isotropically trenched. The trench depth (d) and degree of undercutting (u), along with the resonator width (w) and gap (g), set the interface participation ratios. For example, the MS design was shallowly trenched with a narrow width and gap, while the SA (Si) designs were deeply trenched with narrow (wide) widths and gaps. The MA design was deeply trenched with the resonator mostly suspended.\cite{Woods2019}

% Paragraph 3

All resonators were fabricated on high resistivity 8”~Si(001) substrates ($>$3,500  $\Omega$-cm, Siltronic AG) that were prepared using the RCA clean prior to metal deposition. For the Al, the silicon wafers were also cleaned with an aqueous solution of 1\% hydrofluoric acid (HF) to remove the native oxide prior to deposition. The 750-nm-thick TiN films were deposited in a DC reactive-magnetron sputtering tool (background pressure~$<~2~\times~10^{-8}$~Torr), and the 250-nm-thick Al films were deposited in a molecular-beam epitaxy deposition tool (background pressure~$<~8~\times~10^{-11}$~Torr). The resonator patterns were defined with optical lithography, and the TiN metal was etched with a plasma formed from a combination of $\mathrm{BCl}_{3}$ and $\mathrm{Cl}_{2}$ gasses, whereas the Al metal was etched with a commercial acid etchant. After the resonator patterns were etched, we used an $\mathrm{SF}_{6}$ plasma to isotropically etch the silicon trenches.\cite{Woods2019} Since only the duration of the $\mathrm{SF_{6}}$ plasma etch is varied between resonator designs, we  assume that the loss tangent of each dielectric region in TiN is the same for all geometries. We similarly assume the loss tangent of each dielectric region in Al is the same for all geometries. The TiN device chips without the post-process HF etch came from different wafers than the chips that received the post-process HF etch. Consequentially, the post-etch resonator geometries varied slightly between wafers, which we accounted for by simulating device-specific participation ratios informed by cross-sectional scanning electron micrographs (SEM). All of the Al device chips were taken from the same wafers, and therefore we used the same participation ratios for each set of Al devices. Device chips for the post-process HF etch were etched with a 1\% HF acid solution for approximately 30 seconds to strip the oxides from the surface and then rinsed in deionized water. All device chips were mounted in gold-plated copper packages. The HF-etched chips were loaded into the dilution fridge and pumped down to the millitorr range within 2-3 hours of exposure to atmosphere after the etch to minimize reformation of the native oxides. 

% Paragraph 4 (repositioned)

\begin{figure}
\includegraphics[scale=1]{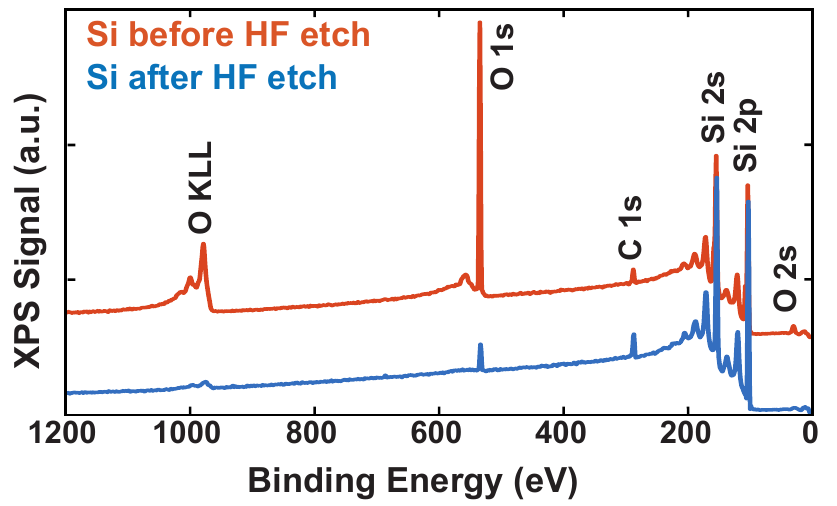}
\caption{\label{fig2} X-ray photoemission spectroscopy (XPS) scan of a representative TiN resonator chip's silicon surface before (orange) and after (blue) the post-process HF treatment. The scans were offset to highlight the reduction in the intensity of the oxygen peaks resulting from the HF etch, indicating the removal of the native silicon oxide.}
\end{figure}

Representative resonator chips were characterized by X-ray photoemission spectroscopy (XPS) before the post-process HF etch (orange line in Fig$.$ \ref{fig2}) and approximately one hour after the post-process HF etch (blue line in Fig$.$ \ref{fig2}). The delay between the post-process HF etch and the XPS scans approximates the time between the post-process HF etch and bringing the devices under vacuum in the dilution fridge. As expected, the oxygen peaks significantly diminished after the HF etch, indicating an effective removal of silicon oxide. In both the Al and TiN metal surfaces, the oxygen peaks were qualitatively unchanged,\cite{Supplemental} suggesting the oxide was reformed faster than the timescale of the experiment.\cite{Dumas1983,Ohya2014} Many previous reports have linked dielectric loss to surface oxides,\cite{Gao2008,Wenner2011,Mueller2019} so we expect that the observed decrease in surface oxides on silicon would result in a decreased substrate-air interface loss tangent after the post-process HF etch.

% Paragraph 5

\begin{figure}
\includegraphics[scale=1]{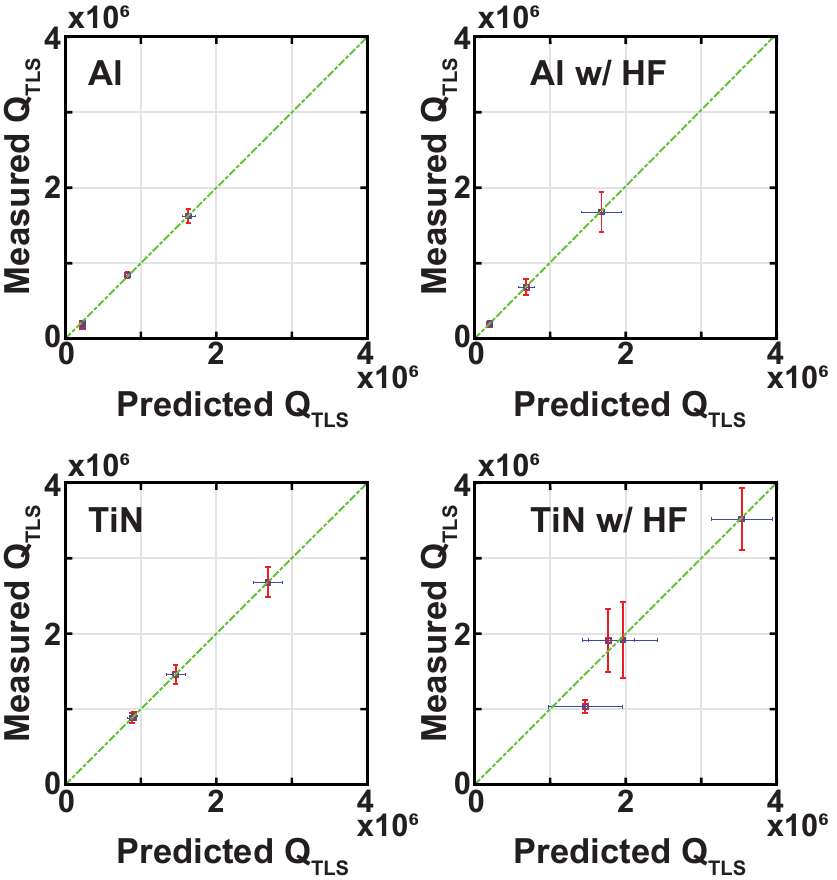}
\caption{\label{fig3}  For each test set, the measured $Q_{\mathrm{TLS}}$ is plotted against the $Q_{\mathrm{TLS}}$ calculated from the participation matrix and the model's loss tangents. Red vertical error bars correspond to the standard error of the measured $Q_{\mathrm{TLS}}$, and blue horizontal error bars correspond to twice the standard deviation of the calculated quality factors. The green line represents perfect agreement between the measured $Q_{\mathrm{TLS}}$ and the predicted $Q_{\mathrm{TLS}}$.}
\end{figure}

To determine the loss tangents, we measured between 10 and 50 resonators of each material and geometry, with and without the post-process HF etch, as a function of circulating microwave power to find the high power ($n_{p}$$\sim$$10^{6}$) and single-photon power ($n_{p}$$\sim$$1$) internal quality factors, $Q_{\mathrm{HP}}$ and $Q_{\mathrm{LP}}$, respectively. We used these to isolate the TLS-limited quality factor, $Q_{\mathrm{TLS}}$,\cite{Calusine2018, Woods2019} defined as
\begin{equation} \label{eqn3}
Q_{\mathrm{TLS}}^{-1} = Q_{\mathrm{LP}}^{-1} - Q_{\mathrm{HP}}^{-1}
\end{equation}
We determined the loss tangent values, $\mathrm{\tan \delta_{r}}$, by applying the SLE Monte Carlo simulation with N = 10,000, using each device set's participation matrix and the $Q_{\mathrm{TLS}}$ values.\cite{Woods2019} The goodness-of-fit of the SLE process is shown in Fig$.$ \ref{fig3}, where we plot the measured $Q_{\mathrm{TLS}}$ against the predicted $Q_{\mathrm{TLS}}$. The \textit{y} values correspond to the mean $Q_{\mathrm{TLS}}$ for each geometry, and the vertical error bars correspond to the standard error of the measured samples. The \textit{x} values correspond to the mean $Q_{\mathrm{TLS}}$ calculated according to Eqn. \ref{eqn1} using the participation matrix and the SLE loss tangents. The horizontal error bars are twice the standard deviation of the calculated quality factors. 

% Paragraph 5.5

The mean and standard deviation of the specific loss tangents determined by the SLE process are given in Table \ref{tab1}, using the dielectric constants and thicknesses in Ref$.$ \onlinecite{Assumptions}. For some regions, e.g., the substrate-air interface for Al, the standard deviation of the loss tangent was much larger than the mean, which we interpreted as the loss tangent being too low to resolve due to the uncertainty in the SLE process given the accessible geometries. Our limited ability to deconvolve the participation of the MS or SA regions relative to the other dielectric regions is one source of uncertainty in the SLE process, which disproportionately affects the metal-substrate (MS) and substrate-air (SA) interfaces compared to the metal-air (MA) interface and the substrate (Si). We also observed greater uncertainty in the SLE process for dielectric regions that minimally affect the total device loss. These effects are apparent in the determination of the TiN substrate-air loss tangent after the post-process HF etch. In the cases where the loss tangent was obscured by the uncertainty, we report the upper bound of the loss tangent for these regions instead of the mean. The metal-substrate and substrate-air upper bounds were set by calculating the highest possible loss tangent consistent with the measured $Q_{\mathrm{TLS}}$ for the MS design and SA design, respectively.\cite{Supplemental}

\begin{table}[!b]
\centering
\caption{Loss tangents for the four dielectric regions by material and process.}
\label{tab1}
\resizebox{\linewidth}{!}{
\renewcommand{\arraystretch}{1.5}
\begin{tabular}{lccccc}
\hline
\multicolumn{1}{c}{}                & \multicolumn{4}{c}{\textbf{Loss tangents}}                                                                                    \\ \hline
\multicolumn{1}{c|}{\textbf{Process}} & \multicolumn{1}{c|}{\textbf{MS ($\times10^{-4}$)}} & \multicolumn{1}{c|}{\textbf{SA ($\times10^{-3}$)}} & \multicolumn{1}{c|}{\textbf{MA ($\times10^{-3}$)}} & \textbf{Si ($\times10^{-7}$)} \\ \hline
\multicolumn{1}{l|}{TiN}              				& \multicolumn{1}{c|}{4.6 $\pm$ 2.4}               	& \multicolumn{1}{c|}{1.7 $\pm$ 0.4}       	& \multicolumn{1}{c|}{3.3 $\pm$ 0.4}          	&		2.6 $\pm$ 0.4                   \\ \hline
\multicolumn{1}{l|}{TiN w/HF}              	& \multicolumn{1}{c|}{2.7 $\pm$ 3.0}               	& \multicolumn{1}{c|}{<1.2}       	& \multicolumn{1}{c|}{3.5 $\pm$ 1.2}          	&		2.8 $\pm$ 0.6            \\ \hline
\multicolumn{1}{l|}{Al}              				& \multicolumn{1}{c|}{<3.2}         		& \multicolumn{1}{c|}{<2.9}         	& \multicolumn{1}{c|}{29.4 $\pm$ 2.9}          	&		2.6 $\pm$ 0.8           \\ \hline
\multicolumn{1}{l|}{Al w/HF}              	& \multicolumn{1}{c|}{<1.3}               	& \multicolumn{1}{c|}{<3.5}       	& \multicolumn{1}{c|}{32.7 $\pm$ 3.6}          	&		1.3 $\pm$ 1.7            \\ \hline
\end{tabular}
}
\end{table}

% Paragraph 6

For both TiN and Al, we found that the loss tangent for the silicon dielectric region is the same within the error bars, as we expected given that (1) the properties of bulk silicon should not change from the fabrication process and (2) the silicon was sourced from the same vendor for all samples. The Si loss tangent we extract is also consistent with other values reported in literature.\cite{Wang2015, Gambetta2017, Woods2019} The most significant material-dependent difference was the loss tangent of the metal-air interface; it is an order of magnitude higher in Al than in TiN. We attribute this to a lossier and thicker aluminum oxide compared to the relatively thin oxide that forms on TiN.\cite{Pappas2011,Gordon2014, Ohya2014, Woods2019} While precise quantitative determination of the metal-substrate and substrate-air loss tangents was not possible in the Al devices, the upper bounds that we set are comparable to their counterparts in TiN without the post-process HF etch.

% Paragraph 7

\begin{figure}
\includegraphics[scale=1]{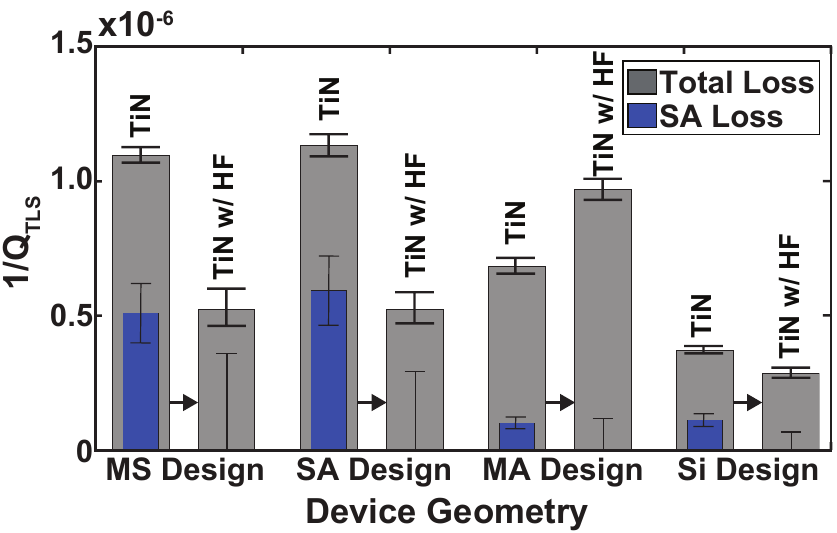}
\caption{\label{fig4} The gray bars represent the measured dielectric loss ($Q_{\mathrm{TLS}}^{-1}$) for each geometry for TiN, with and without the post-process HF etch. The blue bar within each segment represents the predicted loss ascribed to the substrate-air interface. The black arrows indicate the data corresponding to the effect of the post-process HF etch.}
\end{figure}

The only loss tangent that significantly changed due to the post-process HF etch was for the substrate-air interface in TiN, consistent with the expected reduction of silicon oxide at that interface. Although a similar reduction of oxides on the silicon surface occurred in the Al resonator chips from the post-process HF etch, the substrate-air loss tangent was already below the noise floor for Al without the post-process HF etch, and we would not expect to resolve changes to it. In Fig$.$ \ref{fig4}, we plot the effect of the post-process HF etch for each geometry in TiN by comparing the total measured dielectric loss, $Q_{\mathrm{TLS}}^{-1}$, with the calculated dielectric loss from the substrate-air interface. For most geometries, the observed reduction in loss is proportional to the participation ratio of the substrate-air interface. The most significant reductions were observed in the MS and SA designs, corresponding to a decrease of over 50\% in the total measured dielectric loss, and an increase in the overall device performance as determined by the single-photon power internal quality factor, $Q_{LP}$. The $Q_{LP}$ for the MS and SA designs increased from $8.6\times10^{5}$ and $8.0\times10^{5}$ to $1.3\times10^{6}$ and $2.1\times10^{6}$, an increase of 50\% and 160\%, respectively, over the untreated devices. A similar reduction in loss was not observed for the MA design, because the difference in the participation ratio due to the wafer-to-wafer etch variation between the TiN MA designs obscured the impact of the reduced substrate-air loss tangent on total dielectric loss in that geometry. 

% Paragraph 8

In summary, we demonstrated the use of the SLE process developed in Ref$.$ \onlinecite{Woods2019} to quantitatively compare the dielectric loss of superconducting quantum devices made of different materials and fabrication processes. We found that the Al metal-air interface was $\sim$$10\times$ lossier than the TiN metal-air interface. By characterizing the loss at different interfaces, we could strategically target a particularly lossy interface (SA) for improvement. We used a post-process HF etch to reduce the native oxide at that interface, which reduced the substrate-air loss tangent and resulted in more than a $2\times$ increase of the single-photon quality factor compared to untreated devices.

\begin{acknowledgments}
We gratefully acknowledge M. Augeri, P. Baldo, M. Cook, R. Das, M. Hellstrom, V. Iaia, K. Magoon, P. Miller, P. Murphy, B. Osadchy, C. Stull, C. Thoummaraj, and D. Volfson at MIT Lincoln Laboratory for technical assistance. This material is based upon work supported by the Department of Defense under Air Force Contract No. FA8721-05-C-0002 and/or FA8702-15-D-0001. Any opinions, findings, conclusions or recommendations expressed in this material are those of the authors and do not necessarily reflect the views of the Department of Defense.
\end{acknowledgments}

\section*{References}

\appendix*

\bibliography{2821Master}% Produces the bibliography via BibTeX.

\begin{thebibliography}{28}%
\makeatletter
\providecommand \@ifxundefined [1]{%
 \@ifx{#1\undefined}
}%
\providecommand \@ifnum [1]{%
 \ifnum #1\expandafter \@firstoftwo
 \else \expandafter \@secondoftwo
 \fi
}%
\providecommand \@ifx [1]{%
 \ifx #1\expandafter \@firstoftwo
 \else \expandafter \@secondoftwo
 \fi
}%
\providecommand \natexlab [1]{#1}%
\providecommand \enquote  [1]{``#1''}%
\providecommand \bibnamefont  [1]{#1}%
\providecommand \bibfnamefont [1]{#1}%
\providecommand \citenamefont [1]{#1}%
\providecommand \href@noop [0]{\@secondoftwo}%
\providecommand \href [0]{\begingroup \@sanitize@url \@href}%
\providecommand \@href[1]{\@@startlink{#1}\@@href}%
\providecommand \@@href[1]{\endgroup#1\@@endlink}%
\providecommand \@sanitize@url [0]{\catcode `\\12\catcode `\$12\catcode
  `\&12\catcode `\#12\catcode `\^12\catcode `\_12\catcode `\%12\relax}%
\providecommand \@@startlink[1]{}%
\providecommand \@@endlink[0]{}%
\providecommand \url  [0]{\begingroup\@sanitize@url \@url }%
\providecommand \@url [1]{\endgroup\@href {#1}{\urlprefix }}%
\providecommand \urlprefix  [0]{URL }%
\providecommand \Eprint [0]{\href }%
\providecommand \doibase [0]{http://dx.doi.org/}%
\providecommand \selectlanguage [0]{\@gobble}%
\providecommand \bibinfo  [0]{\@secondoftwo}%
\providecommand \bibfield  [0]{\@secondoftwo}%
\providecommand \translation [1]{[#1]}%
\providecommand \BibitemOpen [0]{}%
\providecommand \bibitemStop [0]{}%
\providecommand \bibitemNoStop [0]{.\EOS\space}%
\providecommand \EOS [0]{\spacefactor3000\relax}%
\providecommand \BibitemShut  [1]{\csname bibitem#1\endcsname}%
\let\auto@bib@innerbib\@empty
%</preamble>
\bibitem [{\citenamefont {Martinis}\ \emph {et~al.}(2005)\citenamefont
  {Martinis}, \citenamefont {Cooper}, \citenamefont {McDermott}, \citenamefont
  {Steffen}, \citenamefont {Ansmann}, \citenamefont {Osborn}, \citenamefont
  {Cicak}, \citenamefont {Oh}, \citenamefont {Pappas}, \citenamefont
  {Simmonds},\ and\ \citenamefont {Yu}}]{Martinis2005}%
  \BibitemOpen
  \bibfield  {author} {\bibinfo {author} {\bibfnamefont {J.~M.}\ \bibnamefont
  {Martinis}}, \bibinfo {author} {\bibfnamefont {K.~B.}\ \bibnamefont
  {Cooper}}, \bibinfo {author} {\bibfnamefont {R.}~\bibnamefont {McDermott}},
  \bibinfo {author} {\bibfnamefont {M.}~\bibnamefont {Steffen}}, \bibinfo
  {author} {\bibfnamefont {M.}~\bibnamefont {Ansmann}}, \bibinfo {author}
  {\bibfnamefont {K.~D.}\ \bibnamefont {Osborn}}, \bibinfo {author}
  {\bibfnamefont {K.}~\bibnamefont {Cicak}}, \bibinfo {author} {\bibfnamefont
  {S.}~\bibnamefont {Oh}}, \bibinfo {author} {\bibfnamefont {D.~P.}\
  \bibnamefont {Pappas}}, \bibinfo {author} {\bibfnamefont {R.~W.}\
  \bibnamefont {Simmonds}}, \ and\ \bibinfo {author} {\bibfnamefont {C.~C.}\
  \bibnamefont {Yu}},\ }\bibfield  {title} {\enquote {\bibinfo {title}
  {Decoherence in josephson qubits from dielectric loss},}\ }\href {\doibase
  10.1103/PhysRevLett.95.210503} {\bibfield  {journal} {\bibinfo  {journal}
  {Phys. Rev. Lett.}\ }\textbf {\bibinfo {volume} {95}},\ \bibinfo {pages}
  {210503} (\bibinfo {year} {2005})}\BibitemShut {NoStop}%
\bibitem [{\citenamefont {Gao}\ \emph {et~al.}(2006)\citenamefont {Gao},
  \citenamefont {Mazin}, \citenamefont {Daal}, \citenamefont {Day},
  \citenamefont {LeDuc},\ and\ \citenamefont {Zmuidzinas}}]{Gao2006}%
  \BibitemOpen
  \bibfield  {author} {\bibinfo {author} {\bibfnamefont {J.}~\bibnamefont
  {Gao}}, \bibinfo {author} {\bibfnamefont {B.}~\bibnamefont {Mazin}}, \bibinfo
  {author} {\bibfnamefont {M.}~\bibnamefont {Daal}}, \bibinfo {author}
  {\bibfnamefont {P.}~\bibnamefont {Day}}, \bibinfo {author} {\bibfnamefont
  {H.}~\bibnamefont {LeDuc}}, \ and\ \bibinfo {author} {\bibfnamefont
  {J.}~\bibnamefont {Zmuidzinas}},\ }\bibfield  {title} {\enquote {\bibinfo
  {title} {Power dependence of phase noise in microwave kinetic inductance
  detectors},}\ }in\ \href {\doibase 10.1117/12.672590} {\emph {\bibinfo
  {booktitle} {Millimeter and Submillimeter Detectors and Instrumentation for
  Astronomy {III}}}},\ \bibinfo {editor} {edited by\ \bibinfo {editor}
  {\bibfnamefont {J.}~\bibnamefont {Zmuidzinas}}, \bibinfo {editor}
  {\bibfnamefont {W.~S.}\ \bibnamefont {Holland}}, \bibinfo {editor}
  {\bibfnamefont {S.}~\bibnamefont {Withington}}, \ and\ \bibinfo {editor}
  {\bibfnamefont {W.~D.}\ \bibnamefont {Duncan}}}\ (\bibinfo  {publisher}
  {{SPIE}},\ \bibinfo {year} {2006})\BibitemShut {NoStop}%
\bibitem [{\citenamefont {Gao}\ \emph {et~al.}(2008)\citenamefont {Gao},
  \citenamefont {Daal}, \citenamefont {Martinis}, \citenamefont {Vayonakis},
  \citenamefont {Zmuidzinas}, \citenamefont {Sadoulet}, \citenamefont {Mazin},
  \citenamefont {Day},\ and\ \citenamefont {Leduc}}]{Gao2008}%
  \BibitemOpen
  \bibfield  {author} {\bibinfo {author} {\bibfnamefont {J.}~\bibnamefont
  {Gao}}, \bibinfo {author} {\bibfnamefont {M.}~\bibnamefont {Daal}}, \bibinfo
  {author} {\bibfnamefont {J.~M.}\ \bibnamefont {Martinis}}, \bibinfo {author}
  {\bibfnamefont {A.}~\bibnamefont {Vayonakis}}, \bibinfo {author}
  {\bibfnamefont {J.}~\bibnamefont {Zmuidzinas}}, \bibinfo {author}
  {\bibfnamefont {B.}~\bibnamefont {Sadoulet}}, \bibinfo {author}
  {\bibfnamefont {B.~A.}\ \bibnamefont {Mazin}}, \bibinfo {author}
  {\bibfnamefont {P.~K.}\ \bibnamefont {Day}}, \ and\ \bibinfo {author}
  {\bibfnamefont {H.~G.}\ \bibnamefont {Leduc}},\ }\bibfield  {title} {\enquote
  {\bibinfo {title} {A semiempirical model for two-level system noise in
  superconducting microresonators},}\ }\href {\doibase 10.1063/1.2937855}
  {\bibfield  {journal} {\bibinfo  {journal} {Applied Physics Letters}\
  }\textbf {\bibinfo {volume} {92}},\ \bibinfo {pages} {212504} (\bibinfo
  {year} {2008})}\BibitemShut {NoStop}%
\bibitem [{\citenamefont {O’Connell}\ \emph {et~al.}(2008)\citenamefont
  {O’Connell}, \citenamefont {Ansmann}, \citenamefont {Bialczak},
  \citenamefont {Hofheinz}, \citenamefont {Katz}, \citenamefont {Lucero},
  \citenamefont {McKenney}, \citenamefont {Neeley}, \citenamefont {Wang},
  \citenamefont {Weig}, \citenamefont {Cleland},\ and\ \citenamefont
  {Martinis}}]{OConnell2008}%
  \BibitemOpen
  \bibfield  {author} {\bibinfo {author} {\bibfnamefont {A.~D.}\ \bibnamefont
  {O’Connell}}, \bibinfo {author} {\bibfnamefont {M.}~\bibnamefont
  {Ansmann}}, \bibinfo {author} {\bibfnamefont {R.~C.}\ \bibnamefont
  {Bialczak}}, \bibinfo {author} {\bibfnamefont {M.}~\bibnamefont {Hofheinz}},
  \bibinfo {author} {\bibfnamefont {N.}~\bibnamefont {Katz}}, \bibinfo {author}
  {\bibfnamefont {E.}~\bibnamefont {Lucero}}, \bibinfo {author} {\bibfnamefont
  {C.}~\bibnamefont {McKenney}}, \bibinfo {author} {\bibfnamefont
  {M.}~\bibnamefont {Neeley}}, \bibinfo {author} {\bibfnamefont
  {H.}~\bibnamefont {Wang}}, \bibinfo {author} {\bibfnamefont {E.~M.}\
  \bibnamefont {Weig}}, \bibinfo {author} {\bibfnamefont {A.~N.}\ \bibnamefont
  {Cleland}}, \ and\ \bibinfo {author} {\bibfnamefont {J.~M.}\ \bibnamefont
  {Martinis}},\ }\bibfield  {title} {\enquote {\bibinfo {title} {Microwave
  dielectric loss at single photon energies and millikelvin temperatures},}\
  }\href@noop {} {\bibfield  {journal} {\bibinfo  {journal} {Applied Physics
  Letters}\ }\textbf {\bibinfo {volume} {92}},\ \bibinfo {pages} {112903}
  (\bibinfo {year} {2008})}\BibitemShut {NoStop}%
\bibitem [{\citenamefont {Barends}\ \emph {et~al.}(2008)\citenamefont
  {Barends}, \citenamefont {Hortensius}, \citenamefont {Zijlstra},
  \citenamefont {Baselmans}, \citenamefont {Yates}, \citenamefont {Gao},\ and\
  \citenamefont {Klapwijk}}]{Barends2008}%
  \BibitemOpen
  \bibfield  {author} {\bibinfo {author} {\bibfnamefont {R.}~\bibnamefont
  {Barends}}, \bibinfo {author} {\bibfnamefont {H.~L.}\ \bibnamefont
  {Hortensius}}, \bibinfo {author} {\bibfnamefont {T.}~\bibnamefont
  {Zijlstra}}, \bibinfo {author} {\bibfnamefont {J.~J.~A.}\ \bibnamefont
  {Baselmans}}, \bibinfo {author} {\bibfnamefont {S.~J.~C.}\ \bibnamefont
  {Yates}}, \bibinfo {author} {\bibfnamefont {J.~R.}\ \bibnamefont {Gao}}, \
  and\ \bibinfo {author} {\bibfnamefont {T.~M.}\ \bibnamefont {Klapwijk}},\
  }\bibfield  {title} {\enquote {\bibinfo {title} {Contribution of dielectrics
  to frequency and noise of {NbTiN} superconducting resonators},}\ }\href
  {\doibase 10.1063/1.2937837} {\bibfield  {journal} {\bibinfo  {journal}
  {Applied Physics Letters}\ }\textbf {\bibinfo {volume} {92}},\ \bibinfo
  {pages} {223502} (\bibinfo {year} {2008})}\BibitemShut {NoStop}%
\bibitem [{\citenamefont {Vissers}\ \emph {et~al.}(2010)\citenamefont
  {Vissers}, \citenamefont {Gao}, \citenamefont {Wisbey}, \citenamefont {Hite},
  \citenamefont {Tsuei}, \citenamefont {Corcoles}, \citenamefont {Steffen},\
  and\ \citenamefont {Pappas}}]{Vissers2010}%
  \BibitemOpen
  \bibfield  {author} {\bibinfo {author} {\bibfnamefont {M.~R.}\ \bibnamefont
  {Vissers}}, \bibinfo {author} {\bibfnamefont {J.}~\bibnamefont {Gao}},
  \bibinfo {author} {\bibfnamefont {D.~S.}\ \bibnamefont {Wisbey}}, \bibinfo
  {author} {\bibfnamefont {D.~A.}\ \bibnamefont {Hite}}, \bibinfo {author}
  {\bibfnamefont {C.~C.}\ \bibnamefont {Tsuei}}, \bibinfo {author}
  {\bibfnamefont {A.~D.}\ \bibnamefont {Corcoles}}, \bibinfo {author}
  {\bibfnamefont {M.}~\bibnamefont {Steffen}}, \ and\ \bibinfo {author}
  {\bibfnamefont {D.~P.}\ \bibnamefont {Pappas}},\ }\bibfield  {title}
  {\enquote {\bibinfo {title} {Low loss superconducting titanium nitride
  coplanar waveguide resonators},}\ }\href@noop {} {\bibfield  {journal}
  {\bibinfo  {journal} {Applied Physics Letters}\ }\textbf {\bibinfo {volume}
  {97}},\ \bibinfo {pages} {232509} (\bibinfo {year} {2010})}\BibitemShut
  {NoStop}%
\bibitem [{\citenamefont {Weber}\ \emph {et~al.}(2011)\citenamefont {Weber},
  \citenamefont {Murch}, \citenamefont {Slichter}, \citenamefont {Vijay},\ and\
  \citenamefont {Siddiqi}}]{Weber2011}%
  \BibitemOpen
  \bibfield  {author} {\bibinfo {author} {\bibfnamefont {S.~J.}\ \bibnamefont
  {Weber}}, \bibinfo {author} {\bibfnamefont {K.~W.}\ \bibnamefont {Murch}},
  \bibinfo {author} {\bibfnamefont {D.~H.}\ \bibnamefont {Slichter}}, \bibinfo
  {author} {\bibfnamefont {R.}~\bibnamefont {Vijay}}, \ and\ \bibinfo {author}
  {\bibfnamefont {I.}~\bibnamefont {Siddiqi}},\ }\bibfield  {title} {\enquote
  {\bibinfo {title} {Single crystal silicon capacitors with low microwave loss
  in the single photon regime},}\ }\href {\doibase 10.1063/1.3583449}
  {\bibfield  {journal} {\bibinfo  {journal} {Applied Physics Letters}\
  }\textbf {\bibinfo {volume} {98}},\ \bibinfo {pages} {172510} (\bibinfo
  {year} {2011})}\BibitemShut {NoStop}%
\bibitem [{\citenamefont {Wenner}\ \emph {et~al.}(2011)\citenamefont {Wenner},
  \citenamefont {Barends}, \citenamefont {Bialczak}, \citenamefont {Chen},
  \citenamefont {Kelly}, \citenamefont {Lucero}, \citenamefont {Mariantoni},
  \citenamefont {Megrant}, \citenamefont {O’Malley}, \citenamefont {Sank},
  \citenamefont {Vainsencher}, \citenamefont {Wang}, \citenamefont {White},
  \citenamefont {Yin}, \citenamefont {Zhao}, \citenamefont {Cleland},\ and\
  \citenamefont {Martinis}}]{Wenner2011}%
  \BibitemOpen
  \bibfield  {author} {\bibinfo {author} {\bibfnamefont {J.}~\bibnamefont
  {Wenner}}, \bibinfo {author} {\bibfnamefont {R.}~\bibnamefont {Barends}},
  \bibinfo {author} {\bibfnamefont {R.~C.}\ \bibnamefont {Bialczak}}, \bibinfo
  {author} {\bibfnamefont {Y.}~\bibnamefont {Chen}}, \bibinfo {author}
  {\bibfnamefont {J.}~\bibnamefont {Kelly}}, \bibinfo {author} {\bibfnamefont
  {E.}~\bibnamefont {Lucero}}, \bibinfo {author} {\bibfnamefont
  {M.}~\bibnamefont {Mariantoni}}, \bibinfo {author} {\bibfnamefont
  {A.}~\bibnamefont {Megrant}}, \bibinfo {author} {\bibfnamefont {P.~J.~J.}\
  \bibnamefont {O’Malley}}, \bibinfo {author} {\bibfnamefont
  {D.}~\bibnamefont {Sank}}, \bibinfo {author} {\bibfnamefont {A.}~\bibnamefont
  {Vainsencher}}, \bibinfo {author} {\bibfnamefont {H.}~\bibnamefont {Wang}},
  \bibinfo {author} {\bibfnamefont {T.~C.}\ \bibnamefont {White}}, \bibinfo
  {author} {\bibfnamefont {Y.}~\bibnamefont {Yin}}, \bibinfo {author}
  {\bibfnamefont {J.}~\bibnamefont {Zhao}}, \bibinfo {author} {\bibfnamefont
  {A.~N.}\ \bibnamefont {Cleland}}, \ and\ \bibinfo {author} {\bibfnamefont
  {J.~M.}\ \bibnamefont {Martinis}},\ }\bibfield  {title} {\enquote {\bibinfo
  {title} {Surface loss simulations of superconducting coplanar waveguide
  resonators},}\ }\href@noop {} {\bibfield  {journal} {\bibinfo  {journal}
  {Applied Physics Letters}\ }\textbf {\bibinfo {volume} {99}},\ \bibinfo
  {pages} {113513} (\bibinfo {year} {2011})}\BibitemShut {NoStop}%
\bibitem [{\citenamefont {Chang}\ \emph {et~al.}(2013)\citenamefont {Chang},
  \citenamefont {Vissers}, \citenamefont {Córcoles}, \citenamefont {Sandberg},
  \citenamefont {Gao}, \citenamefont {Abraham}, \citenamefont {Chow},
  \citenamefont {Gambetta}, \citenamefont {Rothwell}, \citenamefont {Keefe},
  \citenamefont {Steffen},\ and\ \citenamefont {Pappas}}]{Chang2013}%
  \BibitemOpen
  \bibfield  {author} {\bibinfo {author} {\bibfnamefont {J.~B.}\ \bibnamefont
  {Chang}}, \bibinfo {author} {\bibfnamefont {M.~R.}\ \bibnamefont {Vissers}},
  \bibinfo {author} {\bibfnamefont {A.~D.}\ \bibnamefont {Córcoles}}, \bibinfo
  {author} {\bibfnamefont {M.}~\bibnamefont {Sandberg}}, \bibinfo {author}
  {\bibfnamefont {J.}~\bibnamefont {Gao}}, \bibinfo {author} {\bibfnamefont
  {D.~W.}\ \bibnamefont {Abraham}}, \bibinfo {author} {\bibfnamefont {J.~M.}\
  \bibnamefont {Chow}}, \bibinfo {author} {\bibfnamefont {J.~M.}\ \bibnamefont
  {Gambetta}}, \bibinfo {author} {\bibfnamefont {M.~B.}\ \bibnamefont
  {Rothwell}}, \bibinfo {author} {\bibfnamefont {G.~A.}\ \bibnamefont {Keefe}},
  \bibinfo {author} {\bibfnamefont {M.}~\bibnamefont {Steffen}}, \ and\
  \bibinfo {author} {\bibfnamefont {D.~P.}\ \bibnamefont {Pappas}},\ }\bibfield
   {title} {\enquote {\bibinfo {title} {Improved superconducting qubit
  coherence using titanium nitride},}\ }\href@noop {} {\bibfield  {journal}
  {\bibinfo  {journal} {Applied Physics Letters}\ }\textbf {\bibinfo {volume}
  {103}},\ \bibinfo {pages} {012602} (\bibinfo {year} {2013})}\BibitemShut
  {NoStop}%
\bibitem [{\citenamefont {Woods}\ \emph {et~al.}(2019)\citenamefont {Woods},
  \citenamefont {Calusine}, \citenamefont {Melville}, \citenamefont {Sevi},
  \citenamefont {Golden}, \citenamefont {Kim}, \citenamefont {Rosenberg},
  \citenamefont {Yoder}, \citenamefont {Dauler},\ and\ \citenamefont
  {Oliver}}]{Woods2019}%
  \BibitemOpen
  \bibfield  {author} {\bibinfo {author} {\bibfnamefont {W.}~\bibnamefont
  {Woods}}, \bibinfo {author} {\bibfnamefont {G.}~\bibnamefont {Calusine}},
  \bibinfo {author} {\bibfnamefont {A.}~\bibnamefont {Melville}}, \bibinfo
  {author} {\bibfnamefont {A.}~\bibnamefont {Sevi}}, \bibinfo {author}
  {\bibfnamefont {E.}~\bibnamefont {Golden}}, \bibinfo {author} {\bibfnamefont
  {D.~K.}\ \bibnamefont {Kim}}, \bibinfo {author} {\bibfnamefont
  {D.}~\bibnamefont {Rosenberg}}, \bibinfo {author} {\bibfnamefont {J.~L.}\
  \bibnamefont {Yoder}}, \bibinfo {author} {\bibfnamefont {E.}~\bibnamefont
  {Dauler}}, \ and\ \bibinfo {author} {\bibfnamefont {W.~D.}\ \bibnamefont
  {Oliver}},\ }\bibfield  {title} {\enquote {\bibinfo {title} {Determining
  interface dielectric losses in superconducting coplanar-waveguide
  resonators},}\ }\href@noop {} {\bibfield  {journal} {\bibinfo  {journal}
  {Physical Review Applied}\ }\textbf {\bibinfo {volume} {12}},\ \bibinfo
  {pages} {014012} (\bibinfo {year} {2019})}\BibitemShut {NoStop}%
\bibitem [{\citenamefont {Wang}\ \emph {et~al.}(2015)\citenamefont {Wang},
  \citenamefont {Axline}, \citenamefont {Gao}, \citenamefont {Brecht},
  \citenamefont {Chu}, \citenamefont {Frunzio}, \citenamefont {Devoret},\ and\
  \citenamefont {Schoelkopf}}]{Wang2015}%
  \BibitemOpen
  \bibfield  {author} {\bibinfo {author} {\bibfnamefont {C.}~\bibnamefont
  {Wang}}, \bibinfo {author} {\bibfnamefont {C.}~\bibnamefont {Axline}},
  \bibinfo {author} {\bibfnamefont {Y.~Y.}\ \bibnamefont {Gao}}, \bibinfo
  {author} {\bibfnamefont {T.}~\bibnamefont {Brecht}}, \bibinfo {author}
  {\bibfnamefont {Y.}~\bibnamefont {Chu}}, \bibinfo {author} {\bibfnamefont
  {L.}~\bibnamefont {Frunzio}}, \bibinfo {author} {\bibfnamefont {M.~H.}\
  \bibnamefont {Devoret}}, \ and\ \bibinfo {author} {\bibfnamefont {R.~J.}\
  \bibnamefont {Schoelkopf}},\ }\bibfield  {title} {\enquote {\bibinfo {title}
  {Surface participation and dielectric loss in superconducting qubits},}\
  }\href@noop {} {\bibfield  {journal} {\bibinfo  {journal} {Applied Physics
  Letters}\ }\textbf {\bibinfo {volume} {107}},\ \bibinfo {pages} {162601}
  (\bibinfo {year} {2015})}\BibitemShut {NoStop}%
\bibitem [{\citenamefont {Gambetta}\ \emph {et~al.}(2017)\citenamefont
  {Gambetta}, \citenamefont {Murray}, \citenamefont {Fung}, \citenamefont
  {McClure}, \citenamefont {Dial}, \citenamefont {Shanks}, \citenamefont
  {Sleight},\ and\ \citenamefont {Steffen}}]{Gambetta2017}%
  \BibitemOpen
  \bibfield  {author} {\bibinfo {author} {\bibfnamefont {J.~M.}\ \bibnamefont
  {Gambetta}}, \bibinfo {author} {\bibfnamefont {C.~E.}\ \bibnamefont
  {Murray}}, \bibinfo {author} {\bibfnamefont {Y.~K.~K.}\ \bibnamefont {Fung}},
  \bibinfo {author} {\bibfnamefont {D.~T.}\ \bibnamefont {McClure}}, \bibinfo
  {author} {\bibfnamefont {O.}~\bibnamefont {Dial}}, \bibinfo {author}
  {\bibfnamefont {W.}~\bibnamefont {Shanks}}, \bibinfo {author} {\bibfnamefont
  {J.~W.}\ \bibnamefont {Sleight}}, \ and\ \bibinfo {author} {\bibfnamefont
  {M.}~\bibnamefont {Steffen}},\ }\bibfield  {title} {\enquote {\bibinfo
  {title} {Investigating surface loss effects in superconducting transmon
  qubits},}\ }\href {\doibase 10.1109/TASC.2016.2629670} {\bibfield  {journal}
  {\bibinfo  {journal} {IEEE Transactions on Applied Superconductivity}\
  }\textbf {\bibinfo {volume} {27}},\ \bibinfo {pages} {1--5} (\bibinfo {year}
  {2017})}\BibitemShut {NoStop}%
\bibitem [{\citenamefont {Barends}\ \emph {et~al.}(2007)\citenamefont
  {Barends}, \citenamefont {Baselmans}, \citenamefont {Hovenier}, \citenamefont
  {Gao}, \citenamefont {Yates}, \citenamefont {Klapwijk},\ and\ \citenamefont
  {Hoevers}}]{Barends2007}%
  \BibitemOpen
  \bibfield  {author} {\bibinfo {author} {\bibfnamefont {R.}~\bibnamefont
  {Barends}}, \bibinfo {author} {\bibfnamefont {J.~J.~A.}\ \bibnamefont
  {Baselmans}}, \bibinfo {author} {\bibfnamefont {J.~N.}\ \bibnamefont
  {Hovenier}}, \bibinfo {author} {\bibfnamefont {J.~R.}\ \bibnamefont {Gao}},
  \bibinfo {author} {\bibfnamefont {S.~J.~C.}\ \bibnamefont {Yates}}, \bibinfo
  {author} {\bibfnamefont {T.~M.}\ \bibnamefont {Klapwijk}}, \ and\ \bibinfo
  {author} {\bibfnamefont {H.~F.~C.}\ \bibnamefont {Hoevers}},\ }\bibfield
  {title} {\enquote {\bibinfo {title} {Niobium and tantalum high q resonators
  forphoton detectors},}\ }\href@noop {} {\bibfield  {journal} {\bibinfo
  {journal} {IEEE Transactions on Applied Superconductivity}\ }\textbf
  {\bibinfo {volume} {17}},\ \bibinfo {pages} {263} (\bibinfo {year}
  {2007})}\BibitemShut {NoStop}%
\bibitem [{\citenamefont {Wang}\ \emph {et~al.}(2009)\citenamefont {Wang},
  \citenamefont {Hofheinz}, \citenamefont {Wenner}, \citenamefont {Ansmann},
  \citenamefont {Bialczak}, \citenamefont {Lenander}, \citenamefont {Lucero},
  \citenamefont {Neeley}, \citenamefont {O'Connell}, \citenamefont {Sank},
  \citenamefont {Weides}, \citenamefont {Cleland},\ and\ \citenamefont
  {Martinis}}]{Wang2009}%
  \BibitemOpen
  \bibfield  {author} {\bibinfo {author} {\bibfnamefont {H.}~\bibnamefont
  {Wang}}, \bibinfo {author} {\bibfnamefont {M.}~\bibnamefont {Hofheinz}},
  \bibinfo {author} {\bibfnamefont {J.}~\bibnamefont {Wenner}}, \bibinfo
  {author} {\bibfnamefont {M.}~\bibnamefont {Ansmann}}, \bibinfo {author}
  {\bibfnamefont {R.~C.}\ \bibnamefont {Bialczak}}, \bibinfo {author}
  {\bibfnamefont {M.}~\bibnamefont {Lenander}}, \bibinfo {author}
  {\bibfnamefont {E.}~\bibnamefont {Lucero}}, \bibinfo {author} {\bibfnamefont
  {M.}~\bibnamefont {Neeley}}, \bibinfo {author} {\bibfnamefont {A.~D.}\
  \bibnamefont {O'Connell}}, \bibinfo {author} {\bibfnamefont {D.}~\bibnamefont
  {Sank}}, \bibinfo {author} {\bibfnamefont {M.}~\bibnamefont {Weides}},
  \bibinfo {author} {\bibfnamefont {A.~N.}\ \bibnamefont {Cleland}}, \ and\
  \bibinfo {author} {\bibfnamefont {J.~M.}\ \bibnamefont {Martinis}},\
  }\bibfield  {title} {\enquote {\bibinfo {title} {Improving the coherence time
  of superconducting coplanar resonators},}\ }\href@noop {} {\bibfield
  {journal} {\bibinfo  {journal} {Applied Physics Letters}\ }\textbf {\bibinfo
  {volume} {95}},\ \bibinfo {pages} {233508} (\bibinfo {year}
  {2009})}\BibitemShut {NoStop}%
\bibitem [{\citenamefont {Sage}\ \emph {et~al.}(2011)\citenamefont {Sage},
  \citenamefont {Bolkhovsky}, \citenamefont {Oliver}, \citenamefont {Turek},\
  and\ \citenamefont {Welander}}]{Sage2011}%
  \BibitemOpen
  \bibfield  {author} {\bibinfo {author} {\bibfnamefont {J.~M.}\ \bibnamefont
  {Sage}}, \bibinfo {author} {\bibfnamefont {V.}~\bibnamefont {Bolkhovsky}},
  \bibinfo {author} {\bibfnamefont {W.~D.}\ \bibnamefont {Oliver}}, \bibinfo
  {author} {\bibfnamefont {B.}~\bibnamefont {Turek}}, \ and\ \bibinfo {author}
  {\bibfnamefont {P.~B.}\ \bibnamefont {Welander}},\ }\bibfield  {title}
  {\enquote {\bibinfo {title} {Study of loss in superconducting coplanar
  waveguide resonators},}\ }\href@noop {} {\bibfield  {journal} {\bibinfo
  {journal} {Journal of Applied Physics}\ }\textbf {\bibinfo {volume} {109}},\
  \bibinfo {pages} {063915} (\bibinfo {year} {2011})}\BibitemShut {NoStop}%
\bibitem [{\citenamefont {Megrant}\ \emph {et~al.}(2018)\citenamefont
  {Megrant}, \citenamefont {Neill}, \citenamefont {Barends}, \citenamefont
  {Chiaro}, \citenamefont {Chen}, \citenamefont {Feigl}, \citenamefont {Kelly},
  \citenamefont {Lucero}, \citenamefont {Mariantoni}, \citenamefont {O'Malley},
  \citenamefont {Sank}, \citenamefont {Vainsencher}, \citenamefont {Wenner},
  \citenamefont {White}, \citenamefont {Yin}, \citenamefont {Zhao},
  \citenamefont {Palmstrom}, \citenamefont {Martinis},\ and\ \citenamefont
  {Cleland}}]{Megrant2012}%
  \BibitemOpen
  \bibfield  {author} {\bibinfo {author} {\bibfnamefont {A.}~\bibnamefont
  {Megrant}}, \bibinfo {author} {\bibfnamefont {C.}~\bibnamefont {Neill}},
  \bibinfo {author} {\bibfnamefont {R.}~\bibnamefont {Barends}}, \bibinfo
  {author} {\bibfnamefont {B.}~\bibnamefont {Chiaro}}, \bibinfo {author}
  {\bibfnamefont {Y.}~\bibnamefont {Chen}}, \bibinfo {author} {\bibfnamefont
  {L.}~\bibnamefont {Feigl}}, \bibinfo {author} {\bibfnamefont
  {J.}~\bibnamefont {Kelly}}, \bibinfo {author} {\bibfnamefont
  {E.}~\bibnamefont {Lucero}}, \bibinfo {author} {\bibfnamefont
  {M.}~\bibnamefont {Mariantoni}}, \bibinfo {author} {\bibfnamefont {P.~J.~J.}\
  \bibnamefont {O'Malley}}, \bibinfo {author} {\bibfnamefont {D.}~\bibnamefont
  {Sank}}, \bibinfo {author} {\bibfnamefont {A.}~\bibnamefont {Vainsencher}},
  \bibinfo {author} {\bibfnamefont {J.}~\bibnamefont {Wenner}}, \bibinfo
  {author} {\bibfnamefont {T.~C.}\ \bibnamefont {White}}, \bibinfo {author}
  {\bibfnamefont {Y.}~\bibnamefont {Yin}}, \bibinfo {author} {\bibfnamefont
  {J.}~\bibnamefont {Zhao}}, \bibinfo {author} {\bibfnamefont {C.~J.}\
  \bibnamefont {Palmstrom}}, \bibinfo {author} {\bibfnamefont {J.~M.}\
  \bibnamefont {Martinis}}, \ and\ \bibinfo {author} {\bibfnamefont {A.~N.}\
  \bibnamefont {Cleland}},\ }\bibfield  {title} {\enquote {\bibinfo {title}
  {Analysis and mitigation of interface losses in trenched superconducting
  coplanar waveguide resonators},}\ }\href@noop {} {\bibfield  {journal}
  {\bibinfo  {journal} {Applied Physics Letters}\ }\textbf {\bibinfo {volume}
  {112}},\ \bibinfo {pages} {062601} (\bibinfo {year} {2018})}\BibitemShut
  {NoStop}%
\bibitem [{\citenamefont {Bruno}\ \emph {et~al.}(2015)\citenamefont {Bruno},
  \citenamefont {de~Lange}, \citenamefont {Asaad}, \citenamefont {van~der
  Enden}, \citenamefont {Langford},\ and\ \citenamefont {DiCarlo}}]{Bruno2015}%
  \BibitemOpen
  \bibfield  {author} {\bibinfo {author} {\bibfnamefont {A.}~\bibnamefont
  {Bruno}}, \bibinfo {author} {\bibfnamefont {G.}~\bibnamefont {de~Lange}},
  \bibinfo {author} {\bibfnamefont {S.}~\bibnamefont {Asaad}}, \bibinfo
  {author} {\bibfnamefont {K.~L.}\ \bibnamefont {van~der Enden}}, \bibinfo
  {author} {\bibfnamefont {N.~K.}\ \bibnamefont {Langford}}, \ and\ \bibinfo
  {author} {\bibfnamefont {L.}~\bibnamefont {DiCarlo}},\ }\bibfield  {title}
  {\enquote {\bibinfo {title} {Reducing intrinsic loss in superconducting
  resonators by surface treatment and deep etching of silicon substrates},}\
  }\href {\doibase 10.1063/1.4919761} {\bibfield  {journal} {\bibinfo
  {journal} {Applied Physics Letters}\ }\textbf {\bibinfo {volume} {106}},\
  \bibinfo {pages} {182601} (\bibinfo {year} {2015})}\BibitemShut {NoStop}%
\bibitem [{\citenamefont {Place}\ \emph {et~al.}(2020)\citenamefont {Place},
  \citenamefont {Rodgers}, \citenamefont {Mundada}, \citenamefont {Smitham},
  \citenamefont {Fitzpatrick}, \citenamefont {Leng}, \citenamefont {Premkumar},
  \citenamefont {Bryon}, \citenamefont {Sussman}, \citenamefont {Cheng},
  \citenamefont {Madhavan}, \citenamefont {Babla}, \citenamefont {Jack},
  \citenamefont {Gyenis}, \citenamefont {Yao}, \citenamefont {Cava},
  \citenamefont {de~Leon},\ and\ \citenamefont {Houck}}]{Place2020}%
  \BibitemOpen
  \bibfield  {author} {\bibinfo {author} {\bibfnamefont {A.~P.~M.}\
  \bibnamefont {Place}}, \bibinfo {author} {\bibfnamefont {L.~V.~H.}\
  \bibnamefont {Rodgers}}, \bibinfo {author} {\bibfnamefont {P.}~\bibnamefont
  {Mundada}}, \bibinfo {author} {\bibfnamefont {B.~M.}\ \bibnamefont
  {Smitham}}, \bibinfo {author} {\bibfnamefont {M.}~\bibnamefont
  {Fitzpatrick}}, \bibinfo {author} {\bibfnamefont {Z.}~\bibnamefont {Leng}},
  \bibinfo {author} {\bibfnamefont {A.}~\bibnamefont {Premkumar}}, \bibinfo
  {author} {\bibfnamefont {J.}~\bibnamefont {Bryon}}, \bibinfo {author}
  {\bibfnamefont {S.}~\bibnamefont {Sussman}}, \bibinfo {author} {\bibfnamefont
  {G.}~\bibnamefont {Cheng}}, \bibinfo {author} {\bibfnamefont
  {T.}~\bibnamefont {Madhavan}}, \bibinfo {author} {\bibfnamefont {H.~K.}\
  \bibnamefont {Babla}}, \bibinfo {author} {\bibfnamefont {B.}~\bibnamefont
  {Jack}}, \bibinfo {author} {\bibfnamefont {A.}~\bibnamefont {Gyenis}},
  \bibinfo {author} {\bibfnamefont {N.}~\bibnamefont {Yao}}, \bibinfo {author}
  {\bibfnamefont {R.~J.}\ \bibnamefont {Cava}}, \bibinfo {author}
  {\bibfnamefont {N.~P.}\ \bibnamefont {de~Leon}}, \ and\ \bibinfo {author}
  {\bibfnamefont {A.~A.}\ \bibnamefont {Houck}},\ }\bibfield  {title} {\enquote
  {\bibinfo {title} {New material platform for superconducting transmonqubits
  with coherence times exceeding 0.3milliseconds},}\ }\href@noop {} {\bibfield
  {journal} {\bibinfo  {journal} {arXiv preprint}\ ,\ \bibinfo {pages}
  {2003.00024}} (\bibinfo {year} {2020})}\BibitemShut {NoStop}%
\bibitem [{\citenamefont {Tsioutsios}\ \emph {et~al.}(2020)\citenamefont
  {Tsioutsios}, \citenamefont {Serniak}, \citenamefont {Diamond}, \citenamefont
  {Sivak}, \citenamefont {Wang}, \citenamefont {Shankar}, \citenamefont
  {Frunzio}, \citenamefont {Schoelkopf},\ and\ \citenamefont
  {Devoret}}]{Tsioutsios2020}%
  \BibitemOpen
  \bibfield  {author} {\bibinfo {author} {\bibfnamefont {I.}~\bibnamefont
  {Tsioutsios}}, \bibinfo {author} {\bibfnamefont {K.}~\bibnamefont {Serniak}},
  \bibinfo {author} {\bibfnamefont {S.}~\bibnamefont {Diamond}}, \bibinfo
  {author} {\bibfnamefont {V.}~\bibnamefont {Sivak}}, \bibinfo {author}
  {\bibfnamefont {Z.}~\bibnamefont {Wang}}, \bibinfo {author} {\bibfnamefont
  {S.}~\bibnamefont {Shankar}}, \bibinfo {author} {\bibfnamefont
  {L.}~\bibnamefont {Frunzio}}, \bibinfo {author} {\bibfnamefont
  {R.}~\bibnamefont {Schoelkopf}}, \ and\ \bibinfo {author} {\bibfnamefont
  {M.}~\bibnamefont {Devoret}},\ }\bibfield  {title} {\enquote {\bibinfo
  {title} {Free-standing silicon shadow masks for transmon qubit
  fabrication},}\ }\href@noop {} {\bibfield  {journal} {\bibinfo  {journal}
  {AIP Advances}\ }\textbf {\bibinfo {volume} {10}},\ \bibinfo {pages} {065120}
  (\bibinfo {year} {2020})}\BibitemShut {NoStop}%
\bibitem [{Sup()}]{Supplemental}%
  \BibitemOpen
  \href@noop {} {}\bibinfo {howpublished} {Please refer to the Supplementary
  Materials for additional details.}\BibitemShut {Stop}%
\bibitem [{Ass()}]{Assumptions}%
  \BibitemOpen
  \href@noop {} {}\bibinfo {note} {We assumed the following thicknesses and
  dielectric constants to generate loss tangents: $t_{\mathrm{MS}}$=2 nm,
  $t_{\mathrm{SA}}$=2 nm, $t_{\mathrm{MA}}$=2 nm,
  $\epsilon_{\mathrm{MS}}$=11.4$\epsilon_{0}$,
  $\epsilon_{\mathrm{SA}}$=4$\epsilon_{0}$, and
  $\epsilon_{\mathrm{MA}}$=10$\epsilon_{0}$}\BibitemShut {NoStop}%
\bibitem [{COM()}]{COMSOL}%
  \BibitemOpen
  \href@noop {} {}\bibinfo {howpublished} {www.comsol.com}\BibitemShut
  {NoStop}%
\bibitem [{\citenamefont {Dumas}\ \emph {et~al.}(1983)\citenamefont {Dumas},
  \citenamefont {Dubarry-Barbe}, \citenamefont {Rivière}, \citenamefont
  {Levy},\ and\ \citenamefont {Corset}}]{Dumas1983}%
  \BibitemOpen
  \bibfield  {author} {\bibinfo {author} {\bibfnamefont {P.}~\bibnamefont
  {Dumas}}, \bibinfo {author} {\bibfnamefont {J.}~\bibnamefont
  {Dubarry-Barbe}}, \bibinfo {author} {\bibfnamefont {D.}~\bibnamefont
  {Rivière}}, \bibinfo {author} {\bibfnamefont {Y.}~\bibnamefont {Levy}}, \
  and\ \bibinfo {author} {\bibfnamefont {J.}~\bibnamefont {Corset}},\
  }\bibfield  {title} {\enquote {\bibinfo {title} {Growth of thin alumina film
  on aluminium at room temperature: a kinetic and spectroscopic study by
  surface plasmon excitation},}\ }\href@noop {} {\bibfield  {journal} {\bibinfo
   {journal} {Journal de Physique Colloques}\ }\textbf {\bibinfo {volume} {44
  (C10)}},\ \bibinfo {pages} {205--208} (\bibinfo {year} {1983})}\BibitemShut
  {NoStop}%
\bibitem [{\citenamefont {Ohya}\ \emph {et~al.}(2014)\citenamefont {Ohya},
  \citenamefont {Chiaro}, \citenamefont {Megrant}, \citenamefont {Neill},
  \citenamefont {Barends}, \citenamefont {Chen}, \citenamefont {Kelly},
  \citenamefont {Low}, \citenamefont {Mutus}, \citenamefont {O’Malley},
  \citenamefont {Roushan}, \citenamefont {Sank}, \citenamefont {Vainsencher},
  \citenamefont {Wenner}, \citenamefont {White}, \citenamefont {Yin},
  \citenamefont {Schultz}, \citenamefont {Palmstrøm}, \citenamefont {Mazin},
  \citenamefont {Cleland},\ and\ \citenamefont {Martinis}}]{Ohya2014}%
  \BibitemOpen
  \bibfield  {author} {\bibinfo {author} {\bibfnamefont {S.}~\bibnamefont
  {Ohya}}, \bibinfo {author} {\bibfnamefont {B.}~\bibnamefont {Chiaro}},
  \bibinfo {author} {\bibfnamefont {A.}~\bibnamefont {Megrant}}, \bibinfo
  {author} {\bibfnamefont {C.}~\bibnamefont {Neill}}, \bibinfo {author}
  {\bibfnamefont {R.}~\bibnamefont {Barends}}, \bibinfo {author} {\bibfnamefont
  {Y.}~\bibnamefont {Chen}}, \bibinfo {author} {\bibfnamefont {J.}~\bibnamefont
  {Kelly}}, \bibinfo {author} {\bibfnamefont {D.}~\bibnamefont {Low}}, \bibinfo
  {author} {\bibfnamefont {J.}~\bibnamefont {Mutus}}, \bibinfo {author}
  {\bibfnamefont {P.~J.~J.}\ \bibnamefont {O’Malley}}, \bibinfo {author}
  {\bibfnamefont {P.}~\bibnamefont {Roushan}}, \bibinfo {author} {\bibfnamefont
  {D.}~\bibnamefont {Sank}}, \bibinfo {author} {\bibfnamefont {A.}~\bibnamefont
  {Vainsencher}}, \bibinfo {author} {\bibfnamefont {J.}~\bibnamefont {Wenner}},
  \bibinfo {author} {\bibfnamefont {T.~C.}\ \bibnamefont {White}}, \bibinfo
  {author} {\bibfnamefont {Y.}~\bibnamefont {Yin}}, \bibinfo {author}
  {\bibfnamefont {B.~D.}\ \bibnamefont {Schultz}}, \bibinfo {author}
  {\bibfnamefont {C.~J.}\ \bibnamefont {Palmstrøm}}, \bibinfo {author}
  {\bibfnamefont {B.~A.}\ \bibnamefont {Mazin}}, \bibinfo {author}
  {\bibfnamefont {A.~N.}\ \bibnamefont {Cleland}}, \ and\ \bibinfo {author}
  {\bibfnamefont {J.~M.}\ \bibnamefont {Martinis}},\ }\bibfield  {title}
  {\enquote {\bibinfo {title} {Room temperature deposition of sputtered tin
  films for superconducting coplanar waveguide resonators},}\ }\href
  {http://stacks.iop.org/0953-2048/27/i=1/a=015009} {\bibfield  {journal}
  {\bibinfo  {journal} {Superconductor Science and Technology}\ }\textbf
  {\bibinfo {volume} {27}},\ \bibinfo {pages} {015009} (\bibinfo {year}
  {2014})}\BibitemShut {NoStop}%
\bibitem [{\citenamefont {Mueller}, \citenamefont {Cole},\ and\ \citenamefont
  {Lisenfeld}(2019)}]{Mueller2019}%
  \BibitemOpen
  \bibfield  {author} {\bibinfo {author} {\bibfnamefont {C.}~\bibnamefont
  {Mueller}}, \bibinfo {author} {\bibfnamefont {J.~H.}\ \bibnamefont {Cole}}, \
  and\ \bibinfo {author} {\bibfnamefont {J.}~\bibnamefont {Lisenfeld}},\
  }\bibfield  {title} {\enquote {\bibinfo {title} {Towards understanding
  two-level-systems in amorphous solids: insights from quantum circuits},}\
  }\href@noop {} {\bibfield  {journal} {\bibinfo  {journal} {Report on Progress
  in Physics}\ }\textbf {\bibinfo {volume} {82}},\ \bibinfo {pages} {124501}
  (\bibinfo {year} {2019})}\BibitemShut {NoStop}%
\bibitem [{\citenamefont {Calusine}\ \emph {et~al.}(2018)\citenamefont
  {Calusine}, \citenamefont {Melville}, \citenamefont {Woods}, \citenamefont
  {Das}, \citenamefont {Stull}, \citenamefont {Bolkhovsky}, \citenamefont
  {Braje}, \citenamefont {Hover}, \citenamefont {Kim}, \citenamefont {Miloshi},
  \citenamefont {Rosenberg}, \citenamefont {Sevi}, \citenamefont {Yoder},
  \citenamefont {Dauler},\ and\ \citenamefont {Oliver}}]{Calusine2018}%
  \BibitemOpen
  \bibfield  {author} {\bibinfo {author} {\bibfnamefont {G.}~\bibnamefont
  {Calusine}}, \bibinfo {author} {\bibfnamefont {A.}~\bibnamefont {Melville}},
  \bibinfo {author} {\bibfnamefont {W.}~\bibnamefont {Woods}}, \bibinfo
  {author} {\bibfnamefont {R.}~\bibnamefont {Das}}, \bibinfo {author}
  {\bibfnamefont {C.}~\bibnamefont {Stull}}, \bibinfo {author} {\bibfnamefont
  {V.}~\bibnamefont {Bolkhovsky}}, \bibinfo {author} {\bibfnamefont
  {D.}~\bibnamefont {Braje}}, \bibinfo {author} {\bibfnamefont
  {D.}~\bibnamefont {Hover}}, \bibinfo {author} {\bibfnamefont {D.~K.}\
  \bibnamefont {Kim}}, \bibinfo {author} {\bibfnamefont {X.}~\bibnamefont
  {Miloshi}}, \bibinfo {author} {\bibfnamefont {D.}~\bibnamefont {Rosenberg}},
  \bibinfo {author} {\bibfnamefont {A.}~\bibnamefont {Sevi}}, \bibinfo {author}
  {\bibfnamefont {J.~L.}\ \bibnamefont {Yoder}}, \bibinfo {author}
  {\bibfnamefont {E.}~\bibnamefont {Dauler}}, \ and\ \bibinfo {author}
  {\bibfnamefont {W.~D.}\ \bibnamefont {Oliver}},\ }\bibfield  {title}
  {\enquote {\bibinfo {title} {Analysis and mitigation of interface losses in
  trenched superconducting coplanar waveguide resonators},}\ }\href@noop {}
  {\bibfield  {journal} {\bibinfo  {journal} {Applied Physics Letters}\
  }\textbf {\bibinfo {volume} {112}},\ \bibinfo {pages} {062601} (\bibinfo
  {year} {2018})}\BibitemShut {NoStop}%
\bibitem [{\citenamefont {Pappas}\ \emph {et~al.}(2011)\citenamefont {Pappas},
  \citenamefont {Vissers}, \citenamefont {Wisbey}, \citenamefont {Kline},\ and\
  \citenamefont {Gao}}]{Pappas2011}%
  \BibitemOpen
  \bibfield  {author} {\bibinfo {author} {\bibfnamefont {D.~P.}\ \bibnamefont
  {Pappas}}, \bibinfo {author} {\bibfnamefont {M.~R.}\ \bibnamefont {Vissers}},
  \bibinfo {author} {\bibfnamefont {D.~S.}\ \bibnamefont {Wisbey}}, \bibinfo
  {author} {\bibfnamefont {J.~S.}\ \bibnamefont {Kline}}, \ and\ \bibinfo
  {author} {\bibfnamefont {J.}~\bibnamefont {Gao}},\ }\bibfield  {title}
  {\enquote {\bibinfo {title} {Two level system loss in superconducting
  microwave resonators},}\ }\href {\doibase 10.1109/tasc.2010.2097578}
  {\bibfield  {journal} {\bibinfo  {journal} {{IEEE} Transactions on Applied
  Superconductivity}\ }\textbf {\bibinfo {volume} {21}},\ \bibinfo {pages}
  {871--874} (\bibinfo {year} {2011})}\BibitemShut {NoStop}%
\bibitem [{\citenamefont {Gordon}\ \emph {et~al.}(2014)\citenamefont {Gordon},
  \citenamefont {Abu-Farsakh}, \citenamefont {Janotti},\ and\ \citenamefont
  {Van~de Walle}}]{Gordon2014}%
  \BibitemOpen
  \bibfield  {author} {\bibinfo {author} {\bibfnamefont {L.}~\bibnamefont
  {Gordon}}, \bibinfo {author} {\bibfnamefont {H.}~\bibnamefont {Abu-Farsakh}},
  \bibinfo {author} {\bibfnamefont {A.}~\bibnamefont {Janotti}}, \ and\
  \bibinfo {author} {\bibfnamefont {C.~G.}\ \bibnamefont {Van~de Walle}},\
  }\bibfield  {title} {\enquote {\bibinfo {title} {Hydrogen bonds in
  $al_{2}o_{3}$ as dissipative two-level systems in superconducting qubits},}\
  }\href@noop {} {\bibfield  {journal} {\bibinfo  {journal} {Nature Scientific
  Reports}\ }\textbf {\bibinfo {volume} {4}},\ \bibinfo {pages} {7590}
  (\bibinfo {year} {2014})}\BibitemShut {NoStop}%
\end{thebibliography}%

\end{document}

\pagenumbering{roman}
\titlespacing*{\section}
{0pt}{0em}{0.25em}
\titlespacing*{\subsection}
{0pt}{10.5em}{0.5em}
\titlespacing*{\subsubsection}
{0pt}{0.5em}{0em}
\titleformat{\chapter}[display]
  {\normalfont\bfseries}
  {\chaptertitlename\ \thechapter}{8pt}{\Huge}
\titlespacing*{\chapter}{0pt}{-50pt}{40pt}

%Start Supplemental Section
\chapter*{Supplemental Information}

\pagenumbering{arabic}

\section*{Participation Matrices}
The participation values used in the main text are given below in the Tables S1-S4.
\vspace{2cm}

\begin{table}[htp]
\centering
\makebox[0pt][c]{\parbox{1.0\textwidth}{%

\begin{minipage}[b]{0.5\hsize}\centering
\renewcommand{\arraystretch}{1.5}
\caption{Participation matrix for TiN \mbox{devices} without post-process HF etch}
\begin{tabular}{lccccc}
\hline
\multicolumn{1}{c|}{}                & \multicolumn{4}{c}{\textbf{Participation (\%)}}                                                                                    \\ \hline
\multicolumn{1}{c|}{\textbf{Design}} & \multicolumn{1}{c|}{\textbf{MS}} & \multicolumn{1}{c|}{\textbf{SA}} & \multicolumn{1}{c|}{\textbf{MA}} & \textbf{Si} \\ \hline
\multicolumn{1}{l|}{MS Design}              				& \multicolumn{1}{c|}{0.274}               	& \multicolumn{1}{c|}{0.147}       	& \multicolumn{1}{c|}{0.017}          	&		86.149                       \\ \hline
\multicolumn{1}{l|}{SA Design}              	& \multicolumn{1}{c|}{0.063}               	& \multicolumn{1}{c|}{0.172}       	& \multicolumn{1}{c|}{0.058}          	&		41.099            \\ \hline
\multicolumn{1}{l|}{MA Design}              				& \multicolumn{1}{c|}{0.014}         		& \multicolumn{1}{c|}{0.029}         	& \multicolumn{1}{c|}{0.084}          	&		10.964            \\ \hline
\multicolumn{1}{l|}{Si Design}              	& \multicolumn{1}{c|}{0.042}               	& \multicolumn{1}{c|}{0.026}       	& \multicolumn{1}{c|}{0.006}          	&		80.5158            \\ \hline
\end{tabular}
	\vspace*{1cm}
    \end{minipage}
\begin{minipage}[b]{0.5\hsize}\centering
\renewcommand{\arraystretch}{1.5}
\caption{Participation matrix for Al \mbox{devices} without post-process HF etch}
\begin{tabular}{lccccc}
\hline
\multicolumn{1}{c|}{}                & \multicolumn{4}{c}{\textbf{Participation (\%)}}                                                                                    \\ \hline
\multicolumn{1}{c|}{\textbf{Design}} & \multicolumn{1}{c|}{\textbf{MS}} & \multicolumn{1}{c|}{\textbf{SA}} & \multicolumn{1}{c|}{\textbf{MA}} & \textbf{Si} \\ \hline
\multicolumn{1}{l|}{MS Design}              				& \multicolumn{1}{c|}{0.297}               	& \multicolumn{1}{c|}{0.156}       	& \multicolumn{1}{c|}{0.017}          	&		87.839                       \\ \hline
\multicolumn{1}{l|}{SA Design}              	& \multicolumn{1}{c|}{0.084}               	& \multicolumn{1}{c|}{0.193}       	& \multicolumn{1}{c|}{0.072}          	&		46.128            \\ \hline
\multicolumn{1}{l|}{MA Design}              				& \multicolumn{1}{c|}{0.014}         		& \multicolumn{1}{c|}{0.041}         	& \multicolumn{1}{c|}{0.076}          	&		15.490            \\ \hline
\multicolumn{1}{l|}{Si Design}              	& \multicolumn{1}{c|}{0.050}               	& \multicolumn{1}{c|}{0.033}       	& \multicolumn{1}{c|}{0.007}          	&		79.543            \\ \hline
\end{tabular}
	\vspace*{1cm}
    \end{minipage}

\begin{minipage}[b]{0.5\hsize}\centering
\renewcommand{\arraystretch}{1.5}
\caption{Participation matrix for TiN \mbox{devices} with post-process HF etch}
\begin{tabular}{lccccc}
\hline
\multicolumn{1}{c|}{}                & \multicolumn{4}{c}{\textbf{Participation (\%)}}                                                                                    \\ \hline
\multicolumn{1}{c|}{\textbf{Design}} & \multicolumn{1}{c|}{\textbf{MS}} & \multicolumn{1}{c|}{\textbf{SA}} & \multicolumn{1}{c|}{\textbf{MA}} & \textbf{Si} \\ \hline
\multicolumn{1}{l|}{MS Design}              				& \multicolumn{1}{c|}{0.271}               	& \multicolumn{1}{c|}{0.147}       	& \multicolumn{1}{c|}{0.018}          	&		85.171                       \\ \hline
\multicolumn{1}{l|}{SA Design}              	& \multicolumn{1}{c|}{0.096}               	& \multicolumn{1}{c|}{0.120}       	& \multicolumn{1}{c|}{0.052}          	&		54.690            \\ \hline
\multicolumn{1}{l|}{MA Design}              				& \multicolumn{1}{c|}{0.020}         		& \multicolumn{1}{c|}{0.047}         	& \multicolumn{1}{c|}{0.092}          	&		14.764            \\ \hline
\multicolumn{1}{l|}{Si Design}              	& \multicolumn{1}{c|}{0.041}               	& \multicolumn{1}{c|}{0.025}       	& \multicolumn{1}{c|}{0.005}          	&		80.249            \\ \hline
\end{tabular}
	\vspace*{1cm}
    \end{minipage}
\begin{minipage}[b]{0.5\hsize}\centering
\renewcommand{\arraystretch}{1.5}
\caption{Participation matrix for Al \mbox{devices} with post-process HF etch}
\begin{tabular}{lccccc}
\hline
\multicolumn{1}{c|}{}                & \multicolumn{4}{c}{\textbf{Participation (\%)}}                                                                                    \\ \hline
\multicolumn{1}{c|}{\textbf{Design}} & \multicolumn{1}{c|}{\textbf{MS}} & \multicolumn{1}{c|}{\textbf{SA}} & \multicolumn{1}{c|}{\textbf{MA}} & \textbf{Si} \\ \hline
\multicolumn{1}{l|}{MS Design}              				& \multicolumn{1}{c|}{0.297}               	& \multicolumn{1}{c|}{0.156}       	& \multicolumn{1}{c|}{0.017}          	&		87.839                       \\ \hline
\multicolumn{1}{l|}{SA Design}              	& \multicolumn{1}{c|}{0.084}               	& \multicolumn{1}{c|}{0.193}       	& \multicolumn{1}{c|}{0.072}          	&		46.128            \\ \hline
\multicolumn{1}{l|}{MA Design}              				& \multicolumn{1}{c|}{0.014}         		& \multicolumn{1}{c|}{0.041}         	& \multicolumn{1}{c|}{0.076}          	&		15.490            \\ \hline
\multicolumn{1}{l|}{Si Design}              	& \multicolumn{1}{c|}{0.050}               	& \multicolumn{1}{c|}{0.033}       	& \multicolumn{1}{c|}{0.007}          	&		79.543            \\ \hline
\end{tabular}
	\vspace*{1cm}
    \end{minipage}%
}}
\end{table}
\pagebreak

\section*{X-ray Photoemission Spectroscopy}
Representative X-ray photoemission spectroscopy (XPS) scans of samples made with the fabrication processes used in the main text are given below in Fig. \ref{fig1}.
\begin{figure}[htp]\centering
\includegraphics[scale=1]{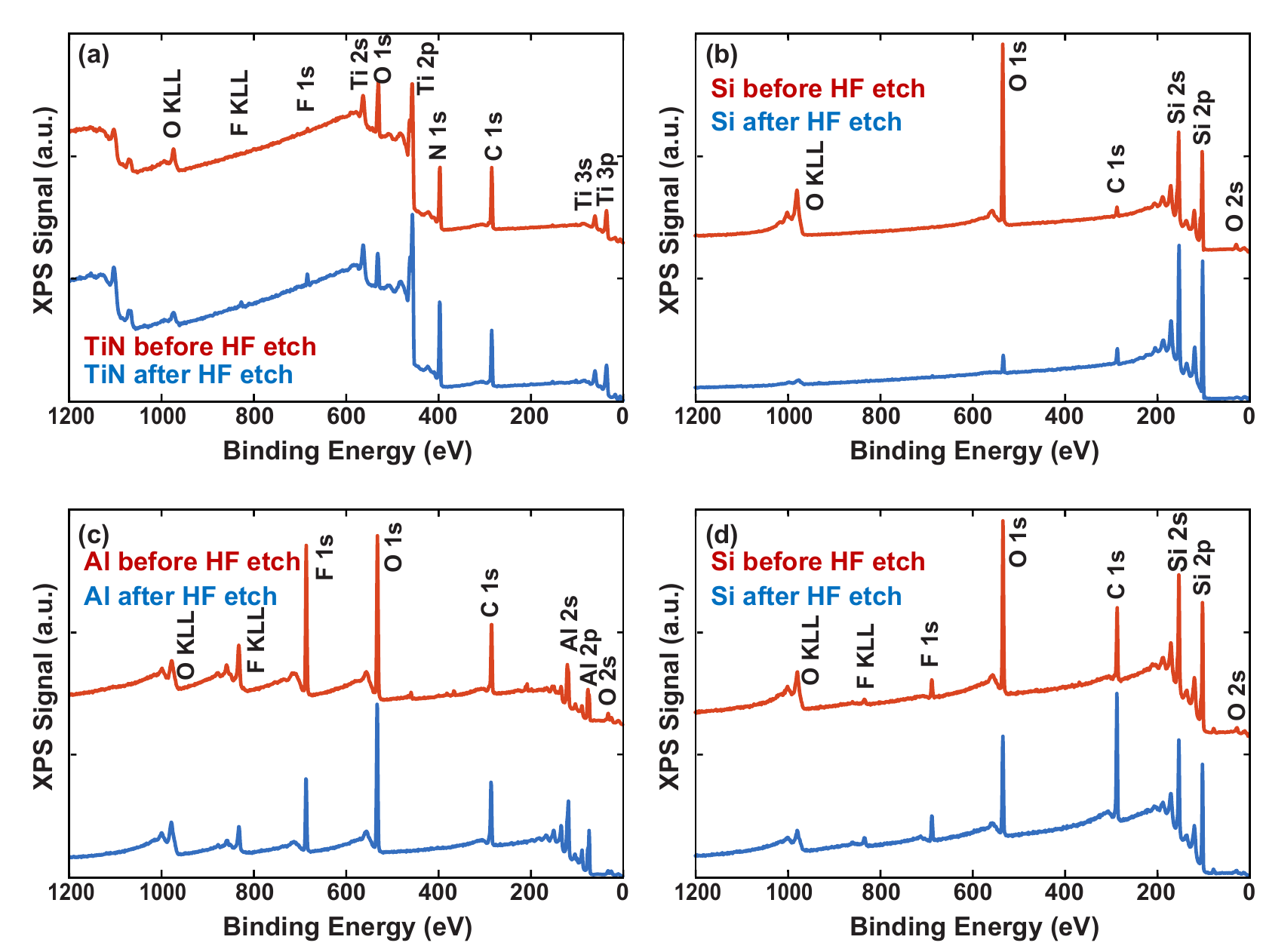}
\caption{\label{fig1} (a) XPS spectroscopy of the titanium nitride surface before (orange) and after (blue) the HF post-process treatment. (b) XPS spectroscopy of the Si surface of the TiN resonator before (orange) and after (blue) the HF post-process treatment. (c) XPS spectroscopy of the Al surface before (orange) and after (blue) the HF post-process treatment. (d) XPS spectroscopy of the Si surface of the Al resonator  before (orange) and after (blue) the HF post-process treatment. The offsets are for better comparison.}
\end{figure}
\pagebreak

\section*{Upper Bounds Calculation}
When the surface-loss extraction process yielded an average loss tangent value of a dielectric region, $\mu_{\tan \delta_{\mathrm{r}}}$, that was much smaller than the standard deviation, $\sigma_{\tan \delta_{\mathrm{r}}}$, we interpreted that as the loss tangent being too low to resolve due to the uncertainty of the model. The algorithm for finding that upper bound is described here. For a given dielectric region, the upper bound was defined as the value for the loss tangent, which, when added to the loss from the minimum value of the other dielectric regions' loss tangents, $\mu_{\tan \delta_{\mathrm{r}}} - 2\sigma_{\tan \delta_{\mathrm{r}}}$, would result in the highest loss observed in the total measured dielectric loss, $\mu_{Q_{\mathrm{TLS}}} - \sigma_{\overline{X}}$, where $\mu_{Q_{\mathrm{TLS}}}$ is the average total measured dielectric loss and $\sigma_{\overline{X}}$ is the standard error of the measured resonators. For example, in the case of the TiN devices with a post-process HF etch, the upper bound for the loss tangent of the substrate-air interface was determined by the equation below.

\begin{equation} \label{eqn1} \centering
\begin{split}
\mathrm{UpperBound}(\tan \delta_{\mathrm{SA}}) = p_{\mathrm{SA}}^{-1}[(\mu_{Q_{\mathrm{TLS,SA Design}}} - \sigma_{\overline{X},\mathrm{SA Design}})^{-1} - p_{\mathrm{MS}}(\mu_{\tan \delta_{\mathrm{MS}}} - 2\sigma_{\tan \delta_{\mathrm{MS}}}) \\ 
- p_{\mathrm{MA}}(\mu_{\tan \delta_{\mathrm{MA}}} - 2\sigma_{\tan \delta_{\mathrm{MA}}}) - p_{\mathrm{Si}}(\mu_{\tan \delta_{\mathrm{Si}}} - 2\sigma_{\tan \delta_{\mathrm{Si}}})]
\end{split}
\end{equation}

\vspace{1cm}
\section*{Aluminum Loss}
\begin{figure}[htp]\centering
\includegraphics[scale=1.5]{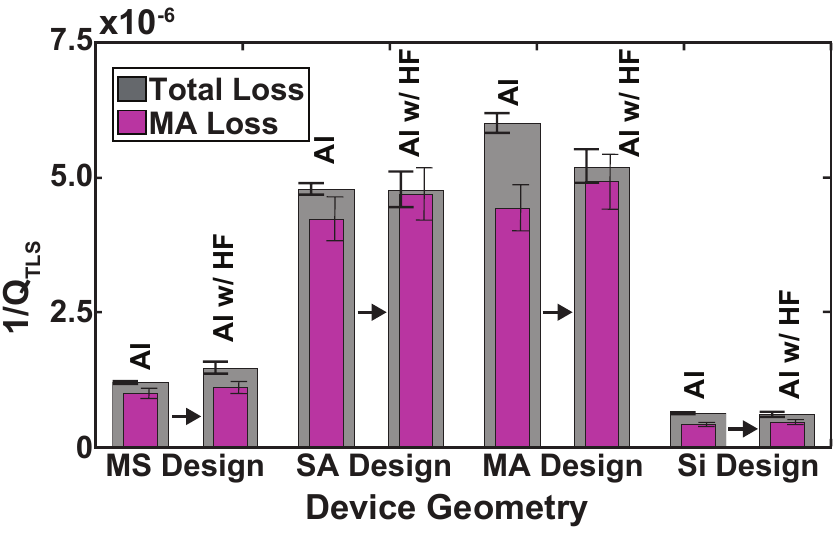}
\caption{\label{fig2} The gray bars represent the measured dielectric loss ($Q_{\mathrm{TLS}}^{-1}$) for each geometry for Al, with and without the post-process HF etch. The pink bar within each segment represents the predicted loss ascribed to the metal-air interface. The black arrows indicate the data corresponding to the effect of the post-process HF etch.}
\end{figure}

In Fig$.$ \ref{fig2}, we plotted the effect of the post-process HF etch for each geometry in Al by comparing the total measured dielectric loss, $Q_{\mathrm{TLS}}^{-1}$, with the calculated dielectric loss from the metal-air interface. The metal-air interface loss dominates in all designs.

